\documentclass[journal]{IEEEtran}

\usepackage{amsmath,amsfonts}
\usepackage{array}
\usepackage[caption=false,font=normalsize,labelfont=sf,textfont=sf]{subfig}
\usepackage{textcomp}
\usepackage{stfloats}
\usepackage{url}
\usepackage{verbatim}
\usepackage{graphicx}
\usepackage{listings}
\usepackage{xcolor}
\usepackage[hidelinks]{hyperref}
\usepackage{orcidlink}
\usepackage{pxfonts}
\usepackage{algorithm}
\usepackage{algpseudocode}
\usepackage{booktabs}
\usepackage{multirow}

\def\BibTeX{{\rm B\kern-.05em{\sc i\kern-.025em b}\kern-.08em
    T\kern-.1667em\lower.7ex\hbox{E}\kern-.125emX}}
\usepackage{balance}

\definecolor{punct}{HTML}{CC3311}
\definecolor{background}{HTML}{EEEEEE}
\definecolor{delim}{HTML}{0077BB}
\definecolor{numb}{HTML}{EE7733}
\definecolor{grey}{HTML}{BBBBBB}
\definecolor{lightgrey}{HTML}{DDDDDD}
\definecolor{lightestgrey}{HTML}{EEEEEE}
\definecolor{keyword}{HTML}{ee7733}

\makeatletter
\lst@InstallKeywords k{attributes}{attributestyle}\slshape{attributestyle}{}ld
\makeatother

\lstdefinelanguage{json}{
    basicstyle=\scriptsize\ttfamily,
    numberstyle=\scriptsize,
    stepnumber=1,
    numbersep=4pt,
    showstringspaces=false,
    breaklines=true,
    frame=lines,
    keywordstyle=\color{keyword}\scshape,
    morekeywords={State, StateMachine, Variable, Guard, Action, Expression, OnTransition, Transition, VariableReference, Event, Property, CollaborativeStateMachine, InvokeAction, CreateAction, AssignAction, DeleteAction, RaiseAction, TimeoutAction, ResetTimeoutAction, MatchAction},
    backgroundcolor=\color{background},
    literate=
     *{0}{{{\color{numb}0}}}{1}
      {1}{{{\color{numb}1}}}{1}
      {2}{{{\color{numb}2}}}{1}
      {3}{{{\color{numb}3}}}{1}
      {4}{{{\color{numb}4}}}{1}
      {5}{{{\color{numb}5}}}{1}
      {6}{{{\color{numb}6}}}{1}
      {7}{{{\color{numb}7}}}{1}
      {8}{{{\color{numb}8}}}{1}
      {9}{{{\color{numb}9}}}{1}
      {:}{{{\color{punct}{:}}}}{1}
      {,}{{{\color{punct}{,}}}}{1}
      {\{}{{{\color{delim}{\{}}}}{1}
      {\}}{{{\color{delim}{\}}}}}{1}
      {[}{{{\color{delim}{[}}}}{1}
      {]}{{{\color{delim}{]}}}}{1}
      {?}{{{\color{delim}{?}}}}{1}
      {+}{{{\color{delim}{+}}}}{1}
      {*}{{{\color{delim}{*}}}}{1}
      {...}{{{\color{punct}{...}}}}{3},
}

\begin{document}

\title{Collaborative State Machines: A Better Programming Model for the Cloud-Edge-IoT Continuum}

\author{Marlon Etheredge\,\orcidlink{0009-0007-3791-9378}, Thomas Fahringer\,\orcidlink{0000-0003-4293-1228}, Felix Erlacher\,\orcidlink{0009-0009-5691-469X}, Elias Kohler\,\orcidlink{0009-0006-3121-6527}, Stefan Pedratscher\,\orcidlink{0000-0002-6164-880X},\\Juan Aznar-Poveda\,\orcidlink{0000-0002-0879-6651}, Nishant Saurabh\,\orcidlink{0000-0002-1926-4693}, Adrien Lebre\,\orcidlink{0000-0002-0305-4130}
\thanks{Marlon Etheredge, Thomas Fahringer, Felix Erlacher, Elias Kohler, Stefan Pedratscher, and Juan Aznar-Poveda are with the Leopold-Franzens-Universität Innsbruck, Austria. E-mail: \{marlon.etheredge, thomas.fahringer, stefan.pedratscher, juan.aznar-poveda\}@uibk.ac.at, \{f.erlacher, elias.kohler\}@student.uibk.ac.at.}
\thanks{Nishant Saurabh is with Utrecht University, Utrecht, The Netherlands. E-mail: n.saurabh@uu.nl.}
\thanks{Adrien Lebre is with Inria, Nantes, France. E-mail: adrien.lebre@inria.fr.}
}

\markboth{}
{Marlon Etheredge, Thomas Fahringer,
\MakeLowercase{\textit{(et al.)}}:
Collaborative State Machines: A Better Programming Model for the Cloud-Edge-IoT Continuum}

\maketitle

\begin{abstract}
The development of Cloud-Edge-IoT applications requires robust programming models. Existing models often struggle to manage the dynamic and stateful nature of these applications effectively. This paper introduces the Collaborative State Machines (CSM) programming model to address these complexities. CSM facilitates the development of reactive, event-driven, and stateful applications targeting the Cloud-Edge-IoT continuum. Applications built with CSM are composed of state machines that collaborate autonomously and can be distributed across different layers of the continuum. Key features of CSM include (i) a sophisticated collaboration mechanism among state machines utilizing events and persistent data; (ii) encapsulation of state through the inherent state of state machines and persistent data; (iii) integration of actions and service invocations within states and state transitions, thereby decoupling complex application logic from compute and data processing services; and (iv) an advanced data model that supports the processing of local, static, and persistent data with defined scope and lifetime. In addition to introducing the CSM programming model, we present a runtime system and a comprehensive evaluation of our approach. This evaluation is based on three use cases: a stress test on a large-scale infrastructure, a surveillance system application, and a complex smart factory scenario, all deployed on the Grid'5000 testbed. Our results demonstrate a 12x increase in throughput through novel language features in the stress test. Compared to Serverless Workflow, a state-of-the-art baseline system, we show a 2.3x improvement in processing time per processed image in a surveillance system use case, a 55x reduction in total processing time for a smart factory use case, and an overall improvement in productivity across these use cases.
\end{abstract}


\section{Introduction}

\IEEEPARstart{T}{he} growing scale and intricacy of Cloud computing have catalyzed the rise of Edge and IoT computing \cite{tankComparativeStudyCloud2023}. This evolution interlinks various application domains, predominantly dependent on sensor data, and necessitates low-latency and high-throughput processing. Cloud-Edge-IoT programming models are instrumental in this context, as they leverage the distinct benefits of both Edge and Cloud environments. At the Edge, they offer low latency, data proximity, and reduced network traffic. At the same time, they harness the robust computing capacity and capabilities, as well as the high availability and reliability of the Cloud. 

Applications targeting the Computing Continuum\footnote{This paper refers to the Cloud-Edge-IoT continuum as the Computing Continuum.} may encompass complex distributed services that use many devices involving large-scale data generation, gathering, storage, and analysis along the Computing Continuum. 

To effectively support application developers targeting the Computing Continuum, innovative programming models and languages must offer advanced support for:

\begin{itemize}
    \item handling event-driven behavior to dynamically respond to system or environmental changes;
    \item maintaining state for data consistency, context awareness, and resilience; and
    \item a sophisticated data model with a well-defined scope and life cycle that goes beyond basic input and output data handling.
\end{itemize}

For instance, traffic management systems in smart cities must process data from numerous IoT sensors to dynamically adjust traffic signals and reduce congestion. Similarly, remote patient monitoring applications in healthcare require continuous state maintenance and context awareness to provide timely alerts and updates. Additionally, industrial IoT applications in smart manufacturing must handle complex event-driven behaviors, such as machinery maintenance alerts and production line adjustments.

\subsection{Limitation of state-of-art approaches}

State-of-the-art programming approaches which are based on workflow \cite{gilExaminingChallengesScientific2007b,tanakaAutomatingEdgetocloudWorkflows2022,smirnovApolloEfficientDistributed2021,amazonwebservicesinc.AWSStepFunctions2024,eismannStateServerlessApplications2022a}, Map-Reduce \cite{deanMapReduceSimplifiedData2008,imaiPerformanceStudyGeoDistributed2018}, and data pipeline \cite{akidauDataflowModelPractical2015,khanSmartDataPlacement2022,carboneApacheFlinkStream2015,raptisSurveyNetworkedData2023} models, are used to implement sequences of conditional tasks that follow a static control- and data-flow pattern. However, they fall short of accommodating data-driven changes in the environment or unexpected events due to their rigid control- and data-flow encoding of all imaginable condition sequences in the application logic. In this, complex applications must be broken down into multiple event-triggered parts that form a single application, resulting in a progressively complex system. Many existing models (e.g., stateless FaaS \cite{eismannStateServerlessApplications2022a}) lack support for stateful computing or require the application developer to manually implement state through centralized data repositories, which implies long latencies. Furthermore, most programming models provide a simple data model, often restricted to input/output data of services, tasks, or pipeline stages. Highly dynamic and event-driven applications can benefit from a more sophisticated data concept that accommodates local, static, and persistent data with different scope and life cycle models. Moreover, existing programming models typically address specific layers of the Computing Continuum and fail to accommodate every layer.

\subsection{Key insights and contributions}

To address the critical limitations inherent in known programming models, we present Collaborative State Machines (CSM) as an innovative programming model tailored for constructing reactive, event-driven, and stateful distributed applications targeting the Computing Continuum. Such applications span various domains, including intelligent systems ('smart anything'), industrial automation, supply chain management, healthcare monitoring, and environmental monitoring. A collaborative state machine involves multiple state machines that can collaborate. Each state machine comprises a collection of states whose transitions are governed by events. Collaboration between these state machines is achieved through event propagation and persistent data. Within each state machine, actions are defined as integral components of states and transitions, facilitating complex behaviors. These actions enable services that can execute computational or data processing tasks.

State machines can be strategically distributed across the layers of the Computing Continuum, allowing the exploitation of distinct layer-specific advantages. Within CSM, state is elevated to a first-class entity, supported by a dual-state concept, encompassing both the inherent state of individual state machines and persistent data, also referred to as shared state \cite{sreekantiCloudburstStatefulFunctionsasaService2020b}. CSM is enriched with an advanced data model for reading and writing data with different contexts (scope and lifetime), including local (created each time a state machine/state is created or entered and destroyed once it is terminated or left), static (preserved value for the lifetime of a state), and persistent (globally available) contexts. This paper makes the following contributions: 

\begin{enumerate}
    \item we introduce the CSM programming model, aimed at the development of event-driven, reactive, and stateful Cloud-Edge-IoT applications;
    \item we introduce the CSM Language (CSML) which supports the CSM programming model;
    \item we showcase the implementation of an open-source CSM runtime system (Cirrina);
    \item we thoroughly evaluate the proposed implementation on the Grid'5000 testbed.
\end{enumerate}

\subsection{Experimental methodology and artifact availability}

The CSML specification and runtime system implementation are open-source projects\footnote{Available from\\\url{https://git.uibk.ac.at/informatik/dps/dps-dc-software/cirrina-project}} supporting an open ecosystem. Our experimental evaluation has been executed on the well-known Grid'5000 testbed for experiment-driven research \cite{balouekAddingVirtualizationCapabilities2013}, providing a realistic testbed spanning several sites and realistic resources, promoting reproducibility. In addition, we provide our use case implementations, which are used to evaluate our runtime system as open-source artifacts. By providing every aspect of our system and evaluation to the public, we strive to make our method and results accessible to the general public. Additionally, we have repeated or executed our experiments over an extended time and gathered results with varying configurations. In our evaluation section, we describe the specific experimentation setup configurations used. Hereby, we adhere to many principles for reproducible evaluation in a Cloud computing context \cite{papadopoulosMethodologicalPrinciplesReproducible2021}.

\subsection{Structure}

The remainder of this paper is structured as follows: Section 2 provides an in-depth exploration of related work; Section 3 introduces the CSM programming model along with the CSM Language; Section 4 introduces our runtime implementation, Cirrina; Section 5 provides our evaluation of three use cases implemented using Cirrina; and Section 6 presents the conclusions and future work.

\section{Related work}

Numerous programming models exist, ranging from workflow-based systems originating from scientific works such as Apollo \cite{smirnovApolloEfficientDistributed2021} and Cloudburst \cite{sreekantiCloudburstStatefulFunctionsasaService2020b} and industry such as AWS Step Functions \cite{amazonwebservicesinc.AWSStepFunctions2024}, Azure Durable Functions \cite{microsoftazureLogicAppService2024}, and Apache Airflow \cite{theapachesoftwarefoundationApacheAirflow2024}. AWS Step Functions workflows are referred to as using 'state machines'; however, they do not provide the intricacies establishing dynamic behavior often associated with conventional state machines and the framework seems to adhere more closely to the commonly used direct-acyclic-graphs (DAG) based models \cite{smirnovApolloEfficientDistributed2021,microsoftazureLogicAppService2024,googlecloudCloudFunctions2024,theapachesoftwarefoundationApacheAirflow2024}. 

Workflow-based systems are primarily centralized, operating within the Cloud layer and require additional services to accommodate additional layers of the Computing Continuum for the development of Cloud-Edge applications such as those provided by public Cloud service providers \cite{microsoftazureIoTEdgeCloud2024,amazonwebservicesinc.IoTEdgeOpen2024,amazonwebservicesinc.SecureIoTGateway2024} and initiatives such as EdgeX \cite{thelinuxfoundationEdgeXFoundryOpen2024}. However, research efforts have been dedicated to the creation of distributed workflow-based systems \cite{smirnovApolloEfficientDistributed2021,tanakaAutomatingEdgetocloudWorkflows2022} targeting the Computing Continuum without emphasizing event-based, reactive, and stateful application behavior. The Pegasus workflow system \cite{tanakaAutomatingEdgetocloudWorkflows2022} has been extended to cater to the Edge-Cloud continuum with a focus on large-scale scientific applications without emphasizing event-based, reactive, and stateful application behavior. To address the limitation of disallowing state transitions within a workflow, TriggerFlow introduced event-driven task triggering by translating workflow descriptions of supported services into an internal representation \cite{arjonaTriggerflowTriggerbasedOrchestration2021}. 

Numerous programming models have been introduced outside workflow-based systems. The actor model \cite{hewittUniversalModularACTOR1973} is a well-known programming paradigm employed in concurrent computing. Programming languages such as Akka \cite{lightbendAkkaBuildConcurrent2024} and Reliable State Machines \cite{mukherjeeReliableStateMachines2019a} provide a lower level of abstraction than our desired high level of abstraction, also seen in other modern Cloud computing description languages such as AWS States Language \cite{amazonwebservicesAmazonStatesLanguage2023}, AFCL \cite{ristovAFCLAbstractFunction2021}, BPMN \cite{chinosiBPMNIntroductionStandard2012}, SCXML \cite{barnettIntroductionSCXML2017}, and XState \cite{statelyXStateJavaScriptState2024}, relying on external services or functions. We recognize the relevance of the actor model and aim to investigate its usage in a future version of our runtime system. ThingML is a code-generation framework that addresses the IoT layer to create heterogeneous application code for different target devices \cite{harrandThingMLLanguageCode2016}. Several widely known programming models specifically address the development of data-processing applications, such as Apache Spark \cite{zahariaApacheSparkUnified2016}, Apache Flink \cite{carboneApacheFlinkStream2015}, and Map-Reduce \cite{deanMapReduceSimplifiedData2008} and also explicitly address the IoT and Edge layers such as FogFlow \cite{chengFogFlowEasyProgramming2018}. Statecharts are supported by frameworks such as Sismic \cite{decanSismicPythonLibrary2020a}, which is a statecharts execution and testing library, and itemis CREATE \cite{itemisagItemisCREATEState}. While itemis CREATE facilitates complex system modeling, it lacks native support for distributed runtime environments. Supported by Red Hat, Serverless Workflow \cite{serverlessworkflowauthorsServerlessWorkflow2024} is recognized as a project within the Cloud Native Computing Foundation (CNCF). It addresses the needs of event-driven and reactive Cloud-Edge-IoT applications with its high-level domain-specific language (DSL) and runtime environments like Apache SonataFlow and Synapse. These tools enable the development of service-oriented architectures across Cloud, Edge, and IoT environments, distinguishing themselves with features such as event-triggered state transitions, dynamic sub-workflow creation, and event generation during transitions.

\section{Collaborative State Machines}

The CSM programming model is built on the concept of distributable state machines that draw inspiration from Harel's statecharts \cite{harelStatechartsVisualFormalism1987a}. State machines emphasize compatibility with statecharts' features. Their primary function lies in exhibiting autonomous, event-driven behavior through state transitions. A collaborative state machine extends beyond individual state machines in statecharts by modeling an application as a set of one or more state machines distributed across the Computing Continuum that collaborate seamlessly through raising and handling events and sharing data.

In a state machine, \emph{actions} can occur during transitions, upon entering or exiting a state, and while present in a state. Actions are crucial as they drive the behavior of state machines, such as triggering events that prompt transitions to subsequent states. Moreover, they facilitate interaction between state machines. Furthermore, actions encompass a variety of essential activities for application functionality, such as invoking services for computation or data manipulation, among other tasks. 

CSM incorporates a comprehensive data model that includes \emph{persistent}, \emph{local}, and \emph{static} data. Persistent data serves as a globally accessible repository, facilitating data sharing across the entire collaborative state machine. Local data is confined to specific state machines or states, enabling efficient data access with a clearly defined scope. Static data has similar efficiency and scope as local data but persists across state transitions.

\begin{figure}
    \centering
    \includegraphics[width=0.8\linewidth]{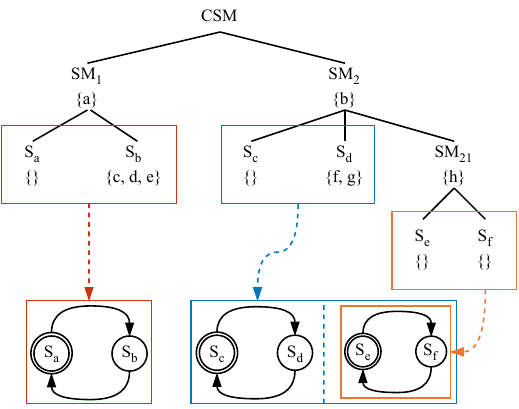}
    \caption{An example of a CSM description containing one nested and two parent state machines as its components. Components can have data represented by $\{a, b, ..., h\}$.}
    \label{fig:example-csm}
\end{figure}

To describe the structure of a collaborative state machine, we introduce the CSM Language (CSML). This description language expresses the control (state transitions) and data flow of a CSM application. In modern workflow description languages, established data serialization languages such as JSON and YAML are commonly used \cite{amazonwebservicesinc.AmazonStatesLanguage2024, ristovAFCLAbstractFunction2021}. CSML adheres to this practice by adopting JSON as a serialization language. Additionally, the language facilitates using expressions defined over data to perform various operations, including evaluating boolean and arithmetic statements and utility functions like random-number generators and list operations. CSML is formally specified in the CSM Specification. 

A CSM description takes the form of a tree structure. At its root is a collaborative state machine, encompassing one or more state machines. A state machine contains one or more states. These fundamental entities --- collaborative state machine, state machines, and states --- are called \emph{components}. Every component can declare local and persistent (globally accessible) data. Figure~\ref{fig:example-csm} illustrates a CSM tree structure. The presented collaborative state machine has two state machines, $\texttt{SM}_1$ and $\texttt{SM}_2$, and a nested state machine denoted as $\texttt{SM}_{21}$. Every state machine has two \emph{atomic states}. The data variables of all components are represented by $\texttt{a}, \texttt{b}, \ldots, \texttt{h}$. The code examples in the following sections will follow the structure of this collaborative state machine.

\begin{lstlisting}[language=json,firstnumber=1,caption={A collaborative state machine construct following the structure of Figure~\ref{fig:example-csm}. The collaborative state machine (\texttt{CSM}) has two state machines and declares a persistent data variable (\texttt{p}).},captionpos=b,label={lst:csm}, mathescape=true, float]
CollaborativeStateMachine{
    name: "CSM",
    memoryMode: "(distributed | shared)",
    stateMachines: [
        StateMachine{ name: "SM$_1$", ...  }*,
        StateMachine{ name: "SM$_2$", ...  }*
    ]
    persistentData?: [ 
        Variable{ name: "p", value: "5 * 5" }* 
    ]
}
\end{lstlisting}

\subsection{Collaborative state machine}

Listing~\ref{lst:csm} shows a collaborative state machine\footnote{In this paper, we use a JSON5-inspired notation for listings. We let \texttt{Construct\{ ... \}} denote a specific CSML language construct. We may include one or more keywords before \texttt{...} for illustrative purposes. Additionally, we use regular expression notation to indicate that one of several options may be chosen (\texttt{a \textbar\ b \textbar\ c}), a keyword is optional (\texttt{?}), a construct may occur zero or more times (\texttt{*}), or one or more times (\texttt{+}).}. The \emph{memory mode} specifies the permitted degree of distribution of state machines across the Computing Continuum. In \emph{distributed} memory mode, the state machines declared within the collaborative state machine may be distributed across multiple resources with distributed memory. In \emph{shared} memory mode, the state machines declared are confined to one resource with private memory.

This collaborative state machine includes two state machines ($\texttt{SM}_1$, $\texttt{SM}_2$). A collaborative state machine may \emph{lexically} declare persistent data. Lexical declaration is performed inherently as part of the description, as opposed to \emph{dynamical} declaration based on actions. Data values may be of any type supported by the runtime system. The collaborative state machine lexically declares one persistent variable (\texttt{p}) globally accessible throughout the collaborative state machine. Its value is derived from an \emph{expression}. Expressions are used to gather data values and may specify a constant or complex operation, such as:

\begin{itemize}
    \item \texttt{5};
    \item \texttt{true};
    \item \texttt{[1, 2, 3]};
    \item \texttt{list\_variable.map(x, x * x)}; or
    \item \texttt{dict\_variable.exists\_one(x, x.success==true)}.
\end{itemize}

The runtime system determines the validity of expression syntax. Variables in scope may be referenced from expressions. In addition to determining the value of a variable as part of a lexical declaration, expressions are frequently used throughout CSML, such as in guard conditions, data assignments, service type invocations (to specify the data values provided to the service), and raising events (to specify the data values provided with the event).

\subsubsection{State machine}

The state machines of a collaborative state machine model event-driven execution behavior, facilitating transitions between atomic states, representing indivisible entities within the state machine reachable via state transitions. Atomic states denote the fundamental states of the state machine. On the other hand, nested state machines operate concurrently within the parent state machine context, allowing for observation and intervention in its execution. These nested structures start simultaneously with their parent. To prevent code duplication, named guards and actions can be declared using the \texttt{guards} and \texttt{actions} keywords, which are referenced by name. Each state machine defines one or more states and optionally includes nested state machines. Within these states, one is designated as the \emph{initial} state, and several \emph{terminal} states may exist. At any given moment, only one state is active within a state machine, while all others remain inactive. Additionally, state machines can declare local data alongside persistent data.

\begin{lstlisting}[language=json,firstnumber=1,caption={A state machine construct. The state machine resembles the structure of $SM_2$ of Figure~\ref{fig:example-csm}, declaring two atomic states ($\texttt{S}_c$ and $\texttt{S}_d$), one nested state machine ($\texttt{SM}_{21}$), and one local data variable (\texttt{b}).},captionpos=b,label={lst:sm}, mathescape=true, float]
StateMachine{
    name: "SM$_2$",
    states: [
        State{ name: "S$_c$", ... }*,
        State{ name: "S$_d$", ... }*,
        StateMachine{ name: "SM$_{21}$", ... }*
    ],
    guards?: [ 
        Guard{ name: "guardA", expression: "b < 100" }*
    ],
    actions?: [
        Action{ name: "actionA", ... }*
    ],
    localData?: [ 
        Variable{ 
            name: "b", 
            value: "[1, 2, 3]" 
        }* 
    ],
    persistentData?: [ Variable{ ... }* ]
}
\end{lstlisting}

Listing~\ref{lst:sm} shows a state machine construct with a name ($\texttt{SM}_2$). In the example, two atomic states ($\texttt{S}_c$, $\texttt{S}_d$) and one nested state machine ($\texttt{SM}_{21}$) are declared. A named guard (\texttt{guardA}) and action (\texttt{actionA}) are declared. These named components can be referenced by name throughout the state machine, reducing code complexity when the guard or action is used multiple times. The guard declares an expression (\texttt{b < 100}) of the form \(E : \mathbb{R}^n \to \{\text{true}, \text{false}\}\), where \(\mathbb{R}^n\) represents the domain of all variables in scope. All guard expressions must be evaluated to true collectively to initiate the transition using the guard. The state machine lexically declares one local variable (\texttt{b}) accessible within the state machine.

\subsubsection{State}

A state within a state machine, which can be initial, terminal, or \emph{intermediate}, becomes active through a state transition. Each state machine must have exactly one initial state. Multiple terminal states can exist, marking the end of the life cycle of the state machine. State transitions are categorized as \emph{always} or \emph{on} transitions. Similar to non-event-driven workflows, \emph{always} transitions occur without events and are used whenever a sequential transition into a subsequent state without event triggering is required. \emph{On} transitions enable the event-driven behavior of the state machine by allowing transitions based on received events. These events can originate from the state machine (\emph{internal} events), other state machines (\emph{external} events), or external sources like applications or devices (\emph{peripheral} events). Guard conditions can be included in the transition declaration to ensure the transition is only taken conditionally.

Actions within a state include \emph{entry}, \emph{exit}, \emph{while}, and \emph{timeout}. \emph{Entry} actions are typically used to initialize or start activities that should occur once as soon as the state is entered, setting up the initial conditions or starting processes necessary for the operation of the state. \emph{Exit} actions, on the other hand, are employed to finalize activities or perform cleanup tasks just before leaving the state, ensuring that any ongoing processes are properly concluded or resources are released. \emph{While} actions, which run continuously while the state is active, are useful for tasks that must be consistently performed throughout the activity of the state, such as monitoring or maintaining certain conditions. Timed actions specified using the \emph{after} keyword, often referred to as \emph{timeout} actions, are used to schedule periodic activities or events within the state, ensuring that specific actions occur at predetermined intervals to manage time-dependent operations. Actions can be either referenced by name or declared in-line. States support the declaration of local, static, and persistent data. \emph{Static} data retains its values between state exits and re-entries, leveraging the re-entry capability of the state. This persistence is particularly useful for maintaining state-specific information that needs to persist across transitions.

\begin{lstlisting}[language=json,firstnumber=1,caption={A state construct. The state resembles the structure of $S_d$ of Figure~\ref{fig:example-csm}. The state declared one static (\texttt{f}) and one local variable (\texttt{g}). The state machine will transition to state $\texttt{S}_c$ if the variable \texttt{g} is evaluated to true.},captionpos=b,label={lst:s}, mathescape=true, float]
State{
    name: "S$_d$",
    (initial | terminal)?: (true|false),
    (entry | exit | while | after)?: [ Action{ ... }* ],
    (on | always)?: [ 
        Transition{ 
            target: "S$_c$",
            guards: [
                Guard{ expression: "g==true" }*
            ],
            ... 
        }* 
    ], 
    staticData?: [
        Variable{ 
            name: "f",
            value: "b.map(x, x * x)"
        }*
    ],
    localData?: [
        Variable{ 
            name: "g",
            value: "f.contains(x, x < 10)"
        }*
    ],
    persistentData?: [ Variable{ ... }* ]
}
\end{lstlisting}

Listing~\ref{lst:s} shows a state construct with a name ({$\texttt{S}_d$}). The state lexically declares one static (\texttt{f}) and one local variable (\texttt{g}). The value of the static variable \texttt{f} is maintained between re-entries of the state $\texttt{S}_d$. Both variables are initialized with complex expressions. Upon entering the state, the local variable \texttt{g} evaluates to \texttt{true} if the static variable \texttt{f}, raised to the power of two, exceeds ten. The static variable \texttt{f} is initialized upon entry into the state.

\subsubsection{Transition}

A state transition within a state enables the transitioning between the states of a state machine. The result of a state transition is exiting the currently active state and entering the new active state. CSML supports \emph{external} transitions, using the \texttt{target} keyword to specify a target state towards which the transition is directed within the same state machine. Optionally, the target state may be omitted, leading to an \emph{internal} transition. In an internal transition, the state machine remains in its current state, bypassing the execution of any actions triggered by entering or exiting the state while still allowing while and timeout actions to continue running. This type of transition ensures the state machine continues uninterrupted, with the possibility of reacting to events by executing specified actions, thereby maintaining continuous operation without exiting the state.

\begin{lstlisting}[language=json,firstnumber=1,caption={A transition construct. When provided as a state transition, upon initiation the state machine will transition to state $\texttt{S}_c$},captionpos=b,label={lst:t}, mathescape=true,float]
Transition{
    target: "S$_c$",
    guards?: [ Guard{ ... }* ],
    actions?: [ Action{ ... }* ]
}
\end{lstlisting}

Listing~\ref{lst:t} illustrates a transition construct; when initiated, this transition leads to entering the state \textit{S$_c$}. During the execution of the transition, its actions are executed. However, the guard conditions must be evaluated to true.

\subsubsection{Actions} \label{par:actions}

Actions in CSML encompass a variety of functionalities, including service invocations, data operations, event raising, and timeouts (referenced in Listings \ref{lst:i}, \ref{lst:d}, \ref{lst:r}, and \ref{lst:to}). These actions are essential for executing critical tasks within states and facilitating interactions both within and between state machines.

\begin{lstlisting}[language=json,firstnumber=1,caption={An invoke action construct. Upon execution, a service with the type \texttt{serviceTypeName} will be invoked.},captionpos=b,label={lst:i}, float]
InvokeAction{
    type: "invoke",
    name: "Name",
    serviceType: "serviceTypeName",
    local?: (true|false),
    input: [ Variable{ ... }* ],
    done: [ Event{ ... }* ],
    properties: [ Property{ ... }* ]
}
\end{lstlisting}

Listing~\ref{lst:i} illustrates a service type invocation action construct used to perform essential application tasks, with these services being executable and accessible through protocols like HTTP or gRPC. In the context of service type invocation, \emph{type} in \emph{service type} refers to the service abstraction of CSML. When invoking a service type, the description does not explicitly define a specific service implementation. Instead, a service type may encompass multiple variations of a particular service, which can be implemented differently; for instance, implementations could use different algorithms, optimize for varying computational resources, or use approximate methods for computational operations to save costs. The decision on the exact implementation to use within a given context is left to the underlying runtime system (see Section \ref{sec:runtime-system}), which can make informed decisions based on the available information of the user or environment (e.g., to optimize for different performance objectives). The \emph{properties} keyword facilitates the specification of hints (e.g., estimated runtime or memory requirements) to the runtime system, which can be used, for instance, to steer the selection of a particular service implementation. Input data for the service type invocation can be provided to the invoked service. To respond to the completion or response of a service invocation, \emph{done} events can be declared. These events allow for handling the outcome of the service execution and triggering subsequent behaviors. In handling these events, the output data of the invoked service is available for storage, passing to subsequent invocations/events or other actions.

Services can be invoked either locally or remotely. Local service invocations occur within the same runtime environment, minimizing potential latency compared to remote invocations. This approach is particularly advantageous for Edge-IoT applications that require access to local resources such as connected cameras, sensors, or the computational capabilities of Edge devices. When specifying a service for local invocation, the runtime system ensures it is executed locally. If not explicitly designated for local invocation, the runtime system decides whether to invoke the service locally or remotely.

\begin{lstlisting}[language=json,firstnumber=1,caption={A dynamic data manipulation action (create, assign or delete) construct. Upon execution, a variable with the name \texttt{v} is respectively created, assigned to, or deleted.},captionpos=b,label={lst:d}, float]
CreateAction{
    type: "create",
    name?: "Name",
    variable: Variable{ name: "v", value: "0" },
    persistent?: (true|false)
}

AssignAction{
    type: "assign",
    name?: "Name",
    variable: VariableReference{ name: "v" },
    value: "v + 1"
}

DeleteAction{
    type: "delete",
    name?: "Name",
    variable: VariableReference{ name: "v" }
}
\end{lstlisting}

In addition to service type invocations, CSML supports various data manipulation actions illustrated by Listing~\ref{lst:d}. These actions enable the dynamic creation of variables as local or persistent data, assigning values to existing variables, and deleting variables. When creating a variable, the \emph{persistent} keyword is used to specify whether the variable should be created persistently. Unlike lexically declaring variables, dynamic data manipulation enables the management of variables throughout the execution of a collaborative state machine.

\begin{lstlisting}[language=json,firstnumber=1,caption={A raise event action construct. Upon execution, an event with the name \texttt{e1} is raised.},captionpos=b,label={lst:r}, float]
RaiseAction{
    type: "raiseEvent",
    name?: "Name",
    event: Event{ 
        name: "e1",
        channel: "(internal | external | global)",
        data: [ Variable{ ... }* ]
    }
}
\end{lstlisting}

Events in the collaborative state machine fall into three categories: \emph{internal}, \emph{external}, and \emph{global}, and are raised by an action as illustrated by Listing~\ref{lst:r}. Internal events are exclusively handled by the state machine that raised them and not seen by other state machines. External events are received by state machines that have subscribed to the events raised by other state machines. This linkage introduces decentralization, allowing state machines to dynamically subscribe to events of interest raised by other state machines. For instance, consider Edge devices expressing interest in events raised by connected IoT devices. The subscription to events from connected IoT devices exemplifies the concept of external events. Global events are not tied to any specific source state machine and are universally seen by every state machine. This universal handling establishes a global communication channel within the application. External sources may raise \emph{peripheral} events handled within the collaborative state machine, allowing the collaborative state machine to respond to events by external software or devices. A raised event may contain data transmitted to the receiver.

\begin{lstlisting}[language=json,firstnumber=1,caption={A reset/timeout action construct. Upon execution, the actions under \emph{actions} are executed every second until the reset timeout action is invoked.},captionpos=b,label={lst:to}, float]
TimeoutAction{
    type: "timeout",
    name?: "TimeoutActionName",
    delay: "1000",
    actions: [ Action{ ... }* ]
}

ResetTimeoutAction{
    type: "resetTimeout",
    name?: "Name",
    action: "TimeoutActionName"
}
\end{lstlisting}

A special type of action, the \emph{timeout} action (illustrated in Listing~\ref{lst:to}), is used in conjunction with the \emph{after} keyword introduced earlier in this section. The timeout action specifies a delay in milliseconds, after which the specified action is executed until reset. The provided action must be a raise event action so subsequent behavior can be triggered upon a raised event. Timeout actions allow the state machine to execute tasks at predefined intervals. Timeouts can be reset based on the \emph{reset timeout} action whenever the execution is no longer required.

\begin{lstlisting}[language=json,firstnumber=1,caption={A match action construct. Upon execution, if the value of variable \texttt{v} is evaluated to be five, the action under \emph{action} is executed.},captionpos=b,label={lst:match}, float]
MatchAction{
    type: "match",
    name?: "Name",
    value: "v",
    cases: [
        {
          case: "5",
          action: Action{ ... }
        }*
    ]
}
\end{lstlisting}

The \emph{match} action is used to allow for conditional action execution. Depending on a \emph{value} expression, one or multiple \emph{case} actions are executed for matching values. In Listing~\ref{lst:match}, the value of a variable \texttt{v} is matched. The corresponding action is executed if a case value in this example \texttt{5} matches the value. 

\subsection{Memory model} \label{sec:memory-mode}

A collaborative state machine operates within two distinct memory modes: shared and distributed. The decision to opt for distributed or shared memory modes may have far-reaching implications. Opting for distribution offers advantages such as heightened parallelism, scalability, replication, and decentralization, although it can lead to increased data transfer times when exchanging data between state machines executed on different resources. Hereafter, we denote the tree representing the CSM description as $CSM$. Moreover, we let $C$ represent any component in $CSM$, $SM$ specifically represent a state machine component, and $S$ specifically represent a state component. In shared memory mode, each $C$ is confined to a computing resource with the same shared memory. Conversely, in distributed memory mode, each $SM$ can be executed on different resources, each equipped with private memory. Thus, these components run on a computing infrastructure with distributed memory. A single state machine operates exclusively in shared memory mode.

The scoping of local data within a collaborative state machine is defined by the shared memory of its components. In shared memory mode, the collaborative state machine $CSM$ executes on a single resource where each component $C \in CSM$ can access local data within its scope $C^\uparrow$, including the component $C$ itself and any of its ancestors up to the root $C_0$. However, in distributed memory mode, the root level of $CSM$ lacks shared memory and cannot declare or access local data. A state machine $SM$ and its nested components $SM_{C}$ always operate in shared memory mode, even within a distributed memory setting. Each state machine $SM$ accesses local data within its scope defined by $SM^\uparrow$. Similarly, each component $SM_{C}$ has its scope denoted by $SM_{C}^\uparrow$, ensuring access to local data is restricted within its scope. This scoping mechanism supports information hiding by restricting data access to ancestors only. For instance, in Figure~\ref{fig:example-csm}, state $S_{e}$ has access to variables $h$ and $b$, state $S_a$ can access variable $a$, and state $S_b$ can access variables $c$, $d$, $e$, and $a$. However, $SM_2$ can only access variable $b$. Sibling components like $S_d$ cannot access data within $SM_{21}$ or $S_c$.

\begin{figure}
    \centering
    \includegraphics[width=0.7\linewidth]{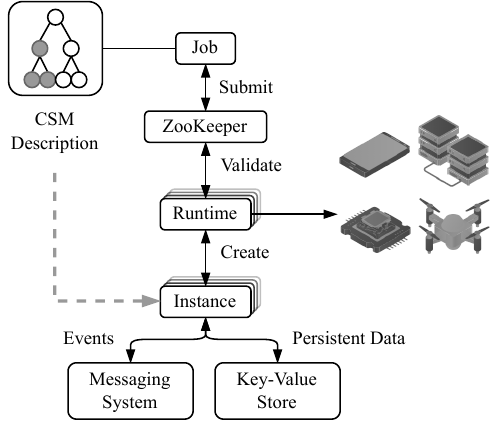}
    \caption{High-level overview of the Cirrina runtime system: Jobs are submitted, processed, and instantiate state machine instances derived from the CSM description.}
    \label{fig:cirrina}
\end{figure}

\subsection{Execution model}

The execution model of CSM is defined according to the execution of an individual state machine. A collaborative state machine containing multiple (nested) state machines will concurrently execute each individual state machine according to the execution model. The execution behavior of a state machine is defined according to the \emph{step algorithm}, taking inspiration from the STATEMATE semantics of statecharts \cite{harelSTATEMATESemanticsStatecharts1996a}. A state machine instance is associated with a \textit{status} $\mu = \langle SM^\uparrow, E, \Theta \rangle$ that captures its current state, including its scope $SM^\uparrow$, input events $E$, and \emph{active configuration} $\Theta$. The active configuration $\Theta$ of the state machine is the set of states $S \in SM$ that are currently \textit{active}. The step algorithm aims to produce a new status according to changes in the state machine's environment to which the state machine responds. The CSM execution model adheres to the following principles:

\begin{enumerate}
    \item changes that occur during a step are effective immediately;
    \item an event is available during one step only and becomes unavailable upon completion of the step where it is handled;
    \item a state machine is considered alive if the active configuration contains no terminal state;
    \item the execution of a step may take time;
    \item actions defined are executed sequentially --- a successive action is executed when the preceding action is completed; and
    \item exactly one state must be active in the active configuration.
\end{enumerate}

(1) Ensures that changes to the status take effect immediately, avoiding non-determinism by committing changes without delay, thus maintaining consistency and predictability in the execution behavior of a state machine. (2) Prevents redundant processing of events within a single step, ensuring that each event is handled exactly once. By making events unavailable after processing in a step, this principle maintains the integrity of event-driven behavior and avoids potential inconsistencies. (3) Defines the lifetime of a state machine based on the absence of terminal states in its active configuration. By considering a state machine alive if it has states to transition through, this principle ensures that execution continues until a terminal state is reached, maintaining the ongoing functionality of the state machine. (4) Acknowledges the potential time consumption during the execution of steps. By recognizing that execution may take time, this principle allows for accommodating delays caused by various factors, ensuring that the execution model remains robust and capable of handling real-world scenarios where computational or external dependencies may affect execution time. (5) Enforces the sequential execution of actions defined within the state machine. By ensuring that each action is executed only after the completion of the preceding action. This principle maintains determinism and prevents race conditions or inconsistencies arising from the concurrent execution of actions, thereby preserving the integrity of state transitions and overall system behavior. (6) Serves to avoid non-determinism, i.e., $|\Theta|=1$, by enforcing that only one state can be active in the active configuration at any given time. Ensuring that the active configuration always contains exactly one active state; this principle prevents ambiguity in state transitions and guarantees predictable behavior within the state machine.

\begin{algorithm}
\caption{Execute a single step.}\label{alg:step}
\hspace*{\algorithmicindent} \textbf{Input:} Status $\mu$ of a state machine. \\
\hspace*{\algorithmicindent} \textbf{Output:} Updated status $\mu$ of the state machine.
\begin{algorithmic}[1]
\Procedure{ExecuteStep}{$\mu$}
\While{$E \neq \emptyset$}
\State $e \gets \text{Pop}(E)$
\State \Call{HandleEvent}{$e, \mu$}
\EndWhile
\EndProcedure
\end{algorithmic}
\end{algorithm}

\begin{algorithm}
\caption{Handle an input event.}\label{alg:handle}
\hspace*{\algorithmicindent} \textbf{Input:} Input event $x$ and status $\mu$. \\
\hspace*{\algorithmicindent} \textbf{Output:} Updated status $\mu$ according to the event.
\begin{algorithmic}[1]
\Procedure{HandleEvent}{$e, \mu$}
\State $\delta \gets \Call{SelectOnTransition}{e, \mu}$
\While{$\delta \neq \emptyset$}
\If{$\delta$ is internal}
\State \Call{Execute}{$\delta$}
\Else
\State \Call{Execute}{cancel, exit, $\delta$, entry, while}
\State \Call{SwitchState}{$\mu$}
\State $\delta \gets \Call{SelectAlwaysTransition}{\mu}$
\EndIf
\EndWhile
\EndProcedure
\end{algorithmic}
\end{algorithm}

The step algorithm consists of two main procedures. Algorithm~\ref{alg:step} processes input events in sequence. Before executing a single step, $E$ is updated with respect to any input events in the environment or collaborative state machine. Peripheral events are added to $E$ before executing a step. Timeout events are treated as events. Algorithm~\ref{alg:handle} is executed for any input event in $E$. An \emph{on} transition is selected (\textit{SelectOnTransition}) iff its source state (the state from which the transition starts) is the active state, its guard conditions evaluate to true, and $e$ is the event that would trigger the transition. An \emph{always} transition is selected iff its source state is the active state, and its guard condition is evaluated to true. The selection of a transition must yield exactly one transition. Conflicts arise whenever one event dictates that the state machine should activate two atomic states. This non-determinism is invalid.

\section{Runtime system} \label{sec:runtime-system}

Our runtime system implementation, Cirrina, is a versatile and extensible CSM runtime system that supports the full range of CSML features. It is fully distributable and accommodates various Cloud service providers, IoT and Edge devices, hardware platforms, and orchestration systems. Additionally, Cirrina introduces novel runtime concepts that enhance CSML capabilities. As an open-source project, it aims to improve accessibility and foster collaboration.

Cirrina, implemented in Java, operates as a distributed CSM runtime deployable across different layers of the Computing Continuum. To ensure efficient execution, Cirrina leverages GraalVM \cite{oracleGraalVM2024} for native compilation on target platforms such as x86-64 and ARM(64). It is designed with distribution in mind, using Apache ZooKeeper \cite{huntZooKeeperWaitfreeCoordination2010} for synchronization when multiple runtimes are deployed. A runtime is the execution environment for \emph{state machine instances}. Within Cirrina, the CSM description in CSML is interpreted to define \emph{state machine classes} along with their states, transitions, and other entities. The runtime can create a state machine instance from a state machine class specified in a CSM description. This approach allows a state machine to be described once and dynamically executed on different resources, avoiding the static application descriptions common in traditional Computing Continuum programming environments.

State machine instances are instantiated with specific instance data, enabling data-driven runtime behavior. For example, a state machine controlling an IoT camera can be provided with its location and ID for communication with other state machine instances. Instances can be configured to handle specific events (binding), facilitating instance-to-instance communication. Extending this example, a state machine instance at the Edge could handle events only from IoT cameras with relevant locations or IDs. Additionally, state machine instances can be supplied with service implementation descriptions, specifying the available service implementations for runtime selection and allowing differentiation based on locality or resource-specific capabilities.

Synchronization between runtimes is managed via ZooKeeper to determine which runtime should create and execute a particular state machine instance. Figure~\ref{fig:cirrina} illustrates this process. A job is submitted to ZooKeeper and watched by all runtimes, containing:

\begin{enumerate}
\item the CSM description;
\item service implementation descriptions, including how to reach the service, the protocol, and location, among other properties;
\item the state machine name to instantiate;
\item instance data;
\item state machine instances to bind events to; and
\item conditions for runtimes to be eligible to run the instance.
\end{enumerate}

Each runtime connected to the ZooKeeper server/cluster receives the job and evaluates the conditions to determine its eligibility to execute the job. Eligible runtimes indicate their ability to run the job via ZooKeeper, and the most suitable runtime is selected to execute the state machine instance. The instance initializes its local data with the instance data contained in the job and its service implementation information with the provided descriptions. The state machine is then executed until termination, communicating with other instances via events and persistent data. Cirrina supports various key-value and messaging systems for events and persistent data, including NATS \cite{natsNATSIo2024} and Kafka \cite{krepsKafkaDistributedMessaging}, and is designed for easy extensibility.

Cirrina supports deployment on multiple orchestration systems, including Kubernetes, Docker Swarm, and AWS ECS. Docker images are available for x86-64 and ARM(64). Cirrina extensively uses OpenTelemetry \cite{opentelemetryauthorsOpenTelemetry2024} to acquire runtime metrics, spans, and other telemetry data that provide insights into the performance of a runtime and CSM application.

\section{Evaluation}

The following section presents three use cases implemented with CSM and evaluated using the Cirrina runtime system. The first use case serves as a stress test for the Cirrina runtime system, demonstrating its capability to invoke services locally and leverage local data efficiently. The second and third use cases involve evaluating two complex applications comparing CSM against a baseline system (Serverless Workflow). All evaluations were conducted on the Grid'5000 testbed.

\subsection{Railway crossing}

\begin{figure}
    \centering
    \includegraphics[width=1.0\linewidth]{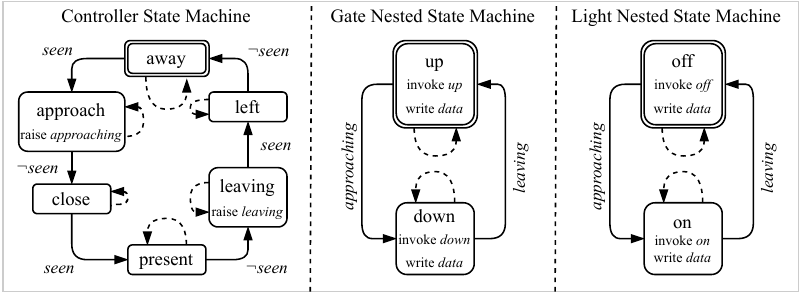}
    \caption{Parent state machine of the railway crossing application and two nested state machines. Dashed edges indicate additional transitions introduced to stress-test the Cirrina runtime.}
    \label{fig:railway}
\end{figure}

Our initial use case aims to stress-test Cirrina in a large-scale distributed environment by increasing the number of events and services invoked, and data writes performed. This scenario centers around a state machine that manages the lights and gates of a railroad crossing, inspired by a classical Petri Net implementation of safety-critical systems by Leveson \cite{levesonSafetyAnalysisUsing1987}. Figure~\ref{fig:railway} illustrates the parent \emph{controller} and two nested \emph{gate} and \emph{light} state machines. The controller state machine primarily responds to events \emph{seen} and \emph{¬seen} from nearby sensors. Event \emph{seen} is triggered when a train is detected by a sensor, while \emph{¬seen} indicates no train. Sensors are placed before, at, and after the railway crossing. By handling sequences of \emph{seen} and \emph{¬seen} events, the state machine monitors the state of approaching and departing trains. The nested state machines control the gates and lights, responding to \emph{approaching} and \emph{leaving} events from the parent state machine. These nested machines invoke services to raise or lower the gates and switch the crossing lights on or off. Additionally, we write data in the form of logs with different payload sizes.

The railway crossing serves as a stress test for Cirrina in a distributed environment, with multiple state machines executed across various resources. The state machine simulates an Edge device managing a railway crossing, with sensors acting as IoT devices. We raise the rate at which events \emph{seen} and \emph{¬seen} occur to increase system load. Additionally, we added self-transitions to every state in the state machine to increase the number of transitions and actions executed, thereby straining the system further by invoking multiple actions (service invocations and data assignments) for each event. We deployed the CSM application using Cirrina on the Grid'5000 testbed. The evaluation setup included:

\begin{itemize}
    \item five sites;
    \item 16 resources across sites;
    \item ten runtimes (two per site); and
    \item 150 state machines (parent and nested).
\end{itemize}

The application was deployed onto the Grid'5000 sites Grenoble, Nantes, Sophia, Nancy, and Rennes. Grid'5000 is organized into geographically distributed sites, each equipped with computing clusters and networking infrastructure, facilitating large-scale experimental research in distributed computing and related fields. We deployed two runtimes per site on nodes from the \emph{dahu}, \emph{econome}, \emph{uvb}, \emph{gros}, and \emph{paranoia} clusters. NATS was employed as a key-value store and messaging system for persistent data storage and event propagation, with one NATS broker per site, forming a five-broker NATS cluster. ZooKeeper was deployed with one server per site, creating a five-server ZooKeeper cluster. This configuration ensured representative measurements across the infrastructure by contacting local NATS brokers and ZooKeeper servers, avoiding unrealistic high latency from inappropriate NATS and ZooKeeper deployments. The gate and light services were implemented in Python using FastAPI, served by the Uvicorn web server, one of the best-performing Python web servers. Services were deployed using the same resources as the runtimes for local services and within the site but outside the runtime resources for remote services. We set the latency to 10ms using \emph{qdisc} to simulate realistic inter-resource latencies. This adjustment compensates for the typically low latency between Grid'5000 nodes, generally under 1ms. The 10ms latency setting is based on common Edge latencies reported in the literature \cite{charyyevLatencyComparisonCloud2020}. Evaluation data was gathered over two hours, with the event rate increasing from low to high every five minutes. We conducted the experiment with two configurations: one using remote service invocations and persistent data (\emph{Remote / Persistent}), and the other using local service invocations and local data (\emph{Local / Local}). In addition to examining the performance of Cirrina, we aim to highlight the importance of integrating a sophisticated data model and advanced service invocations into the programming model. The data presented hereafter represents five-minute intervals, with the total experiment time divided into five-minute segments, indicating the number of repetitions of the experiment. The data presented herein was acquired using OpenTelemetry integrated into our runtime system.

\begin{figure}
    \centering
    \includegraphics[width=0.8\linewidth]{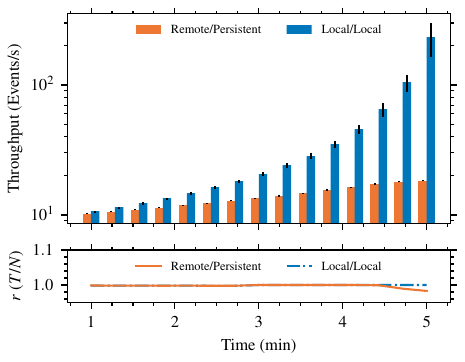}
    \caption{Throughput comparison of two configurations in the railway crossing experiment. The bottom chart illustrates the ratio of throughput ($T$) to generated events ($N$).}
    \label{fig:railway-events}
\end{figure}

Figure~\ref{fig:railway-events} displays the throughput in events per second and the number of generated events for the two configurations. The results demonstrate a significant 12x improvement in throughput with CSM when it invokes services and stores data locally, compared to an approach that uses remote service invocations and persistent data storage. At the maximum event rate, configuration \emph{Remote / Persistent} sees the ratio $r (T/N)$ with $T$ as throughput and $N$ as the number of generated events drop below 1, resulting in event queue accumulation and performance degradation (the system cannot process all generated events in time). In contrast, configuration \emph{Local / Local} maintains a ratio of 1, ensuring real-time throughput relative to generated events. Furthermore, we observe a significantly lower number of handled events per time unit, demonstrating that CSM utilizing local services and local data notably enhances performance compared to traditional approaches.

\begin{figure}
    \centering
    \includegraphics[width=0.8\linewidth]{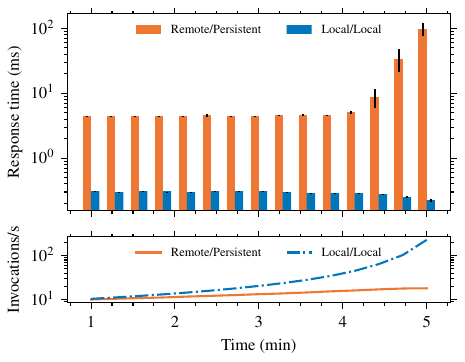}
    \caption{Comparison of response times between two configurations in the railway crossing experiment. The bottom chart depicts the number of invocations performed per second over the same time interval.}
    \label{fig:railway-response-time}
\end{figure}

\begin{figure}
    \centering
    \includegraphics[width=0.8\linewidth]{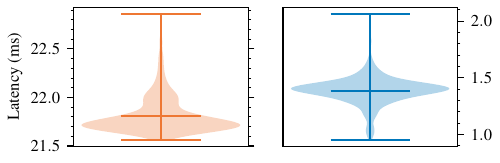}
    \caption{Service invocation latencies of the two configurations, \emph{Persistent / Remote} left, and \emph{Local / Local} right.}
    \label{fig:railway-invoke-latency}
\end{figure}

\begin{figure}
    \centering
    \includegraphics[width=0.8\linewidth]{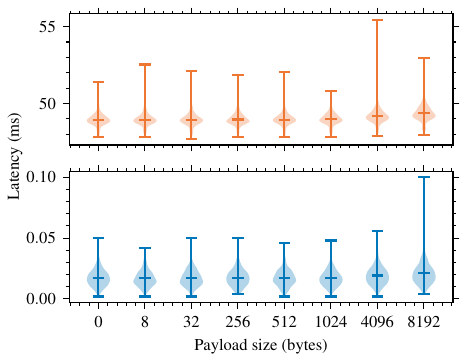}
    \caption{Write latencies of the two configurations, \emph{Persistent / Remote} top, and \emph{Local / Local} bottom, with different payload sizes.}
    \label{fig:railway-write-latency}
\end{figure}

Figure~\ref{fig:railway-response-time} displays response times\footnote{We consistently observed that the web service shows decreasing latency at low latency values as the request rate per second increases up to a certain threshold. Beyond this threshold, latency values begin to rise. This pattern was also evident when the service was tested in isolation and with different service implementations.} for handling events across both configurations, which is measured as the time between handling an event and the time when subsequently executed actions have been completed. Configuration \emph{Remote / Persistent} exhibits higher response times due to the latency associated with remote service invocations, as illustrated in Figure~\ref{fig:railway-invoke-latency}, and data assignment, as shown in Figure~\ref{fig:railway-write-latency}. This latency limits the maximum number of service invocations per second. Conversely, Configuration \emph{Local / Local} demonstrates lower response times and improved throughput, underscoring the performance benefits of local data storage and service invocation, particularly beneficial for Edge and IoT deployments. These results highlight the capability of CSM to enhance performance by leveraging local resources while maintaining the flexibility to utilize remote services and persistent data as needed.

\subsection{Surveillance system}

\begin{figure}
    \centering
    \includegraphics[width=1.0\linewidth]{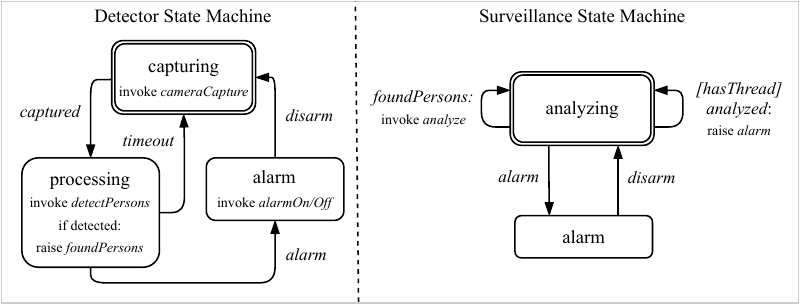}
    \caption{Two state machines forming the surveillance system application. The \emph{Detector} state machine is deployed in the Edge layer, and the \emph{Surveillance} state machine is deployed in the Cloud layer.}
    \label{fig:surveillance}
\end{figure}

Our second use case involves a distributed Cloud-Edge-IoT surveillance system, represented by the collaborative state machine shown in Figure~\ref{fig:surveillance}. The surveillance setup comprises an IoT layer with cameras and alarms, capturing images for processing at the Edge layer. At the Edge layer, images are pre-processed to filter out irrelevant images that do not contain persons, preventing unnecessary analysis, as the primary function of the application is to trigger alerts upon detecting unknown individuals in captured images. By conducting this lightweight pre-processing, the Edge layer ensures that only relevant images are forwarded to the Cloud layer for more computationally intensive analysis. Image capturing was configured to occur every 500ms.

In the CSM implementation, we deploy two state machines: the \emph{Detector} on Edge runtimes and the \emph{Surveillance} on Cloud runtimes. Events coordinate the activation of both state machines:

\begin{enumerate}
\item in the \emph{capturing} state, an image is captured from an IoT device;
\item in the \emph{processing} state, lightweight pre-processing (person detection) occurs;
\item iff persons are detected (\emph{foundPersons} event), Cloud analysis is initiated in the \emph{analyzing} state;
\item upon detecting a threat (unknown person), a global alarm is triggered;
\item if no persons are detected, the \emph{Detector} state machine times out in the \emph{processing} state and initiates a new capture; and
\item an alarm can be disarmed via the \emph{disarm} event, once triggered.
\end{enumerate}

The application depicted in Figure~\ref{fig:surveillance} was implemented using CSM and Serverless Workflow. Since the message processor and monitoring system operate independently from the main functionality, we opted not to include these components in the Serverless Workflow implementation; however, they are incorporated in the CSM implementation. Serverless Workflow \cite{serverlessworkflowauthorsServerlessWorkflow2024} is a cutting-edge system with a domain-specific language designed for defining workflows. It handles dynamic, event-driven applications and supports deployment in distributed environments. We deployed both applications on the Grid'5000 testbed. The evaluation setup included:

\begin{itemize}
    \item three sites (two Edge, one Cloud);
    \item 10 resources across sites;
    \item six runtimes (two per site); and
    \item 18 workflows (Serverless Workflow), 12 state machines (CSM).
\end{itemize}

Deployments spanned Grid'5000 sites in Grenoble, Nantes, and Rennes, utilizing Cirrina as the CSM runtime and Apache Sonataflow as the Serverless Workflow runtime system. We evaluated multiple Serverless Workflow runtimes, including Synapse, and found Apache Sonataflow to offer the most advanced runtime implementation. The experimental setup mirrored that of our first use case, with an additional simulated latency of 20ms introduced to the Cloud resources and services to simulate a realistic latency to reach the Cloud \cite{charyyevLatencyComparisonCloud2020}. To optimize performance, the Serverless Workflow application underwent rigorous profiling and benchmarking before evaluation.

In our evaluation, we found the processing time of a single image for the Serverless Workflow implementation of the surveillance system application to have a mean of 763ms and a standard deviation of 38ms. The equivalent CSM implementation was found to have a mean of 335ms and a standard deviation of 49ms. Processing time was measured from image capture to alarm triggering, covering the IoT (capture), Edge (pre-processing), and Cloud (analysis) layers of the Serverless Workflow and CSM implementations. This data was collected over two hours of execution. These observations found that the CSM implementation has an average processing time 2.3x lower than the Serverless Workflow implementation per processed image for an equivalent application and services. On average, the processing time of a single image using Serverless Workflow lies above the capturing rate of 500ms, indicating that the accumulation of captured images could occur, resulting in progressive performance degradation. One of the key advantages of CSM is its ability to continue execution without frequent re-executions of workflows, unlike a system such as Serverless Workflow. While Serverless Workflow can manage long-running workflows where events drive state transitions, which is an improvement upon competing systems, CSM applications benefit from state machines addressing specific application concerns, such as Edge-specific person detection or Cloud-specific face analysis. This approach eliminates the need to re-execute parts of the application upon certain events, significantly reducing application overhead. Furthermore, the event semantics of CSM enable precise targeting of events to the relevant state machines, avoiding the costly broadcasting of events intended for specific tasks. These results highlight the significant performance advantage of CSM over Serverless Workflow. This underscores the potential of CSM in enhancing Cloud-Edge-IoT applications by reducing processing times, thereby improving responsiveness and overall system effectiveness.

\subsection{Smart factory}

\begin{figure}
    \centering
    \includegraphics[width=1.0\linewidth]{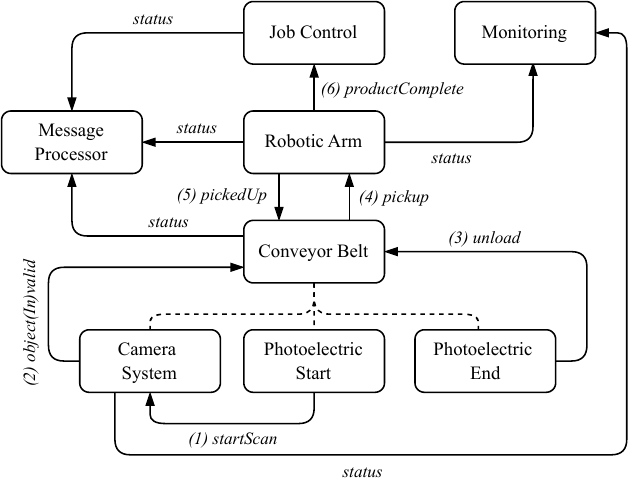}
    \caption{Components of the smart factory application. Solid edges indicate raised and handled events, while dashed edges represent composition.}
    \label{fig:smart-factory}
\end{figure}

The third use case we present is a smart factory application. This application utilizes the full Computing Continuum by modeling a production line that includes conveyor belts, robotic arms, camera systems within the Edge-IoT domain, and monitoring and assembly management in the Cloud domain. Figure~\ref{fig:smart-factory} illustrates the resulting application, represented by state machines using CSM, depicting the interactions between state machines through events raised and handled. The application features a conveyor belt equipped with a camera system and two photoelectric sensors --- one at the start and one at the end of the conveyor belt. The process begins with placing an object on the conveyor belt (1). The camera system then determines whether the object is valid (2), which dictates whether the conveyor belt continues moving or stops. When the object reaches the end of the conveyor belt (3), a signal is sent to a robotic arm (4), indicating that the object is ready to be picked up and processed. The conveyor belt cycle repeats after the robotic arm processes the object (5). Once sufficient processing has been performed and a product is completed (6), it undergoes further processing. Additional status events are sent at various stages throughout this process.

We implemented the application shown in Figure~\ref{fig:smart-factory} using CSM and Serverless Workflow. Serverless Workflow provides a state-of-the-art programming model capable of providing a solid baseline due to its capabilities to accommodate event-driven and reactive applications. We deployed both applications on the Grid'5000 testbed. The setup included:

\begin{itemize}
    \item three sites (two Edge, one Cloud);
    \item 10 resources across sites;
    \item six runtimes (two per site); and
    \item 36 - 156 workflows (Serverless Workflow), 36 - 156 state machines (CSM).
\end{itemize} 

Both applications were deployed across Grid'5000 sites in Grenoble, Nantes, and Rennes. Cirrina served as the CSM runtime, while Apache Sonataflow was chosen as the Serverless Workflow runtime system. Each site hosted two runtimes deployed on nodes from the \emph{dahu}, \emph{econome}, \emph{paravance}, and \emph{paranoia} clusters. For Cirrina, NATS and ZooKeeper were deployed similarly to the first two experiments, with both implementations invoking the same services. Efforts were made to optimize the Serverless Workflow application for performance by profiling and benchmarking it before evaluation, and we iteratively developed and profiled the application to ensure its performance in the evaluation.

\begin{figure}
    \centering
    \includegraphics[width=0.8\linewidth]{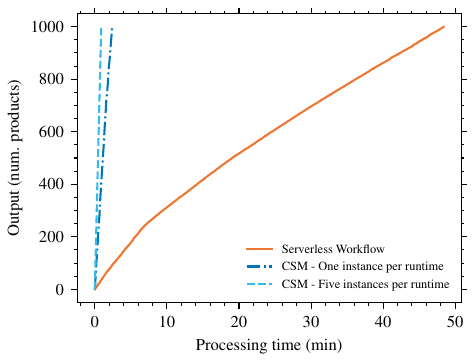}
    \caption{Production times comparison for the smart factory use case between Serverless Workflow and CSM implementations. Two CSM configurations are shown: one with a single instance per runtime and another with five instances per runtime.}
    \label{fig:smart-factory-processing-time}
\end{figure}

Figure~\ref{fig:smart-factory-processing-time} illustrates the production times for the Serverless Workflow and CSM implementations in assembling 1000 products under the highest event rate achievable. Through experimentation, we determined these event rates to be 200ms for Apache SonataFlow and 10ms for Cirrina. The 200ms limit reflects the workflow-based system's implicit requirement for a higher volume of events due to the necessary subdivision of workflows to accommodate certain application behaviors.  In addition to investigating scaling with increased event rates, we assessed how both systems scale with the number of deployed instances (conveyor belt including its components and robotic arm). Increasing the number of instances within the smart factory is expected to reduce production time, as each application instance contributes to product output per unit of time.

We observed that Apache SonataFlow, constrained by its event rate, faces limitations when scaling with the number of applications. In Figure~\ref{fig:smart-factory-processing-time}, the Serverless Workflow application required 2904 seconds to assemble 1000 products, while the CSM application, with one instance per runtime, achieved this in 144 seconds. Increasing the number of instances per runtime to five reduced production time to 52 seconds, representing a 2.8x reduction facilitated by the ability of Cirrina to coordinate resource collaboration. Comparing the most efficient CSM implementation shows a 56x improvement over an equivalent Serverless Workflow implementation of the smart factory application. This improvement can be attributed to the capability of Cirrina to handle significantly higher event rates due to its advanced event semantics and the reduced overhead inherent in CSM applications.

\subsection{Productivity}

\begin{table}[]
     \caption{Productivity metrics of the surveillance system and smart factory use cases, comparing the equivalent Serverless Workflow and CSM implementations.}
    \label{tab:productivity}
    \resizebox{0.48\textwidth}{!}{
    \begin{tabular}{lcccc}
    \hline
    \multirow{2}{*}{Metric / Use case}                                               & \multicolumn{2}{c}{Surveillance system}  & \multicolumn{2}{c}{Smart factory}  \\ \cline{2-5} 
                                                                                     & Serverless Workflow         & CSM        & Serverless Workflow      & CSM     \\ \hline
    Lines of code                                                                    & 344                         & 244        & 1264                     & 922     \\
    Number of states                                                                 & 16                          & 5          & 42                       & 22      \\
    \begin{tabular}[c]{@{}l@{}}Number of Workflows \\ or state machines\end{tabular} & 3                           & 2          & 6                        & 6       \\ \hline
    \end{tabular}}
\end{table}

Table~\ref{tab:productivity} presents productivity metrics for the surveillance system and smart factory use cases, illustrating metrics for both the Serverless Workflow and CSM implementations. The table shows that the CSM implementations generally have fewer lines of code, states, and workflows or state machines compared to equivalent Serverless Workflow implementations, leading to an overall increase in productivity when using CSM instead of Serverless Workflow.

\section{Conclusions and Future Work}

Collaborative State Machines (CSM) represents a novel programming model to develop reactive, event-driven, and stateful applications for the Computing Continuum. A collaborative state machine may consist of multiple state machines, each of which can model and implement highly dynamic and event-driven application logic that can run on any number of resources across the Computing Continuum. Interaction within and across state machines is supported through events and persistent data. To manage the state of an application, system, environment or individual components of them, or individual components of both, a dual-state concept is offered by design. Actions and service (computational and data processing tasks) invocations offer fine-grain control of highly dynamic and complex application logic. An advanced data model with scope and lifetime across state machines enables sophisticated data-driven application logic. The CSM programming model facilitates the development of intricate Cloud-Edge-IoT applications through its dedicated language, CSML. An efficient runtime system, Cirrina, has been developed and thoroughly evaluated. Cirrina is a robust implementation of CSM, enhancing its capabilities with novel runtime features such as state machine instantiation.

Our evaluation, conducted on the Grid'5000 testbed using a stress test CSM application, demonstrates that CSM, in conjunction with the Cirrina runtime system, manages high throughput and achieves low processing times for event responses. Further enhancements in throughput are achievable through leveraging local service invocations and data. In specific applications like a Cloud-Edge-IoT surveillance system implemented using CSM and Serverless Workflow, we observed a notable 2.3x reduction in processing time per image. Similarly, in a Cloud-Edge-IoT smart factory application scenario, CSM compared with Serverless Workflow yielded an impressive 56x reduction in total processing time. Additionally, these implementations demonstrated an overall improvement in productivity. These results underscore the effectiveness of CSM and the Cirrina runtime system in optimizing performance across diverse Cloud-Edge-IoT environments, promising substantial advancements in application efficiency and responsiveness.

In our future work, we will explore the opportunities presented by the significant computational power of state machines. Our focus will additionally include investigating advanced scheduling techniques for CSM applications, encompassing both the instances managed by the runtimes and the strategic allocation and placement of the runtimes themselves within the continuum.

The CSML specifications, Cirrina, as well as the use case implementations, are published online: \url{https://git.uibk.ac.at/informatik/dps/dps-dc-software/cirrina-project}

\section*{Acknowledgements}

Experiments presented in this paper were carried out using the Grid'5000 testbed, supported by a scientific interest group hosted by Inria and including CNRS, RENATER, and several Universities as well as other organizations (see \url{https://www.grid5000.fr}). 

\bibliographystyle{IEEEtran}
\bibliography{bibliography}

\begin{IEEEbiography}[{\includegraphics
[width=1in,height=1.25in,clip,
keepaspectratio]{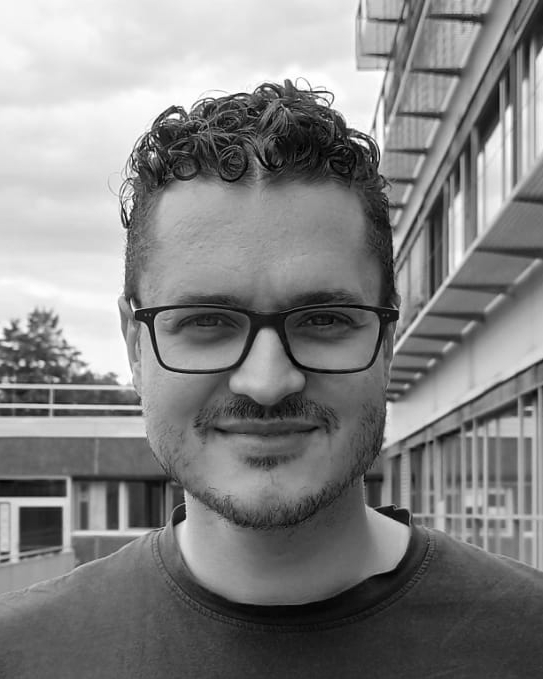}}]
{Marlon Etheredge} 
received his M.Sc in Artificial Intelligence from the University of Utrecht, The Netherlands, in 2016. He is a Ph.D. student in the Distributed and Parallel Systems Group at the University of Innsbruck, Austria, and his main research interests include programming paradigms and runtime systems in the Cloud-Edge-IoT continuum.
\end{IEEEbiography}

\begin{IEEEbiography}[{\includegraphics
[width=1in,height=1.25in,clip,
keepaspectratio]{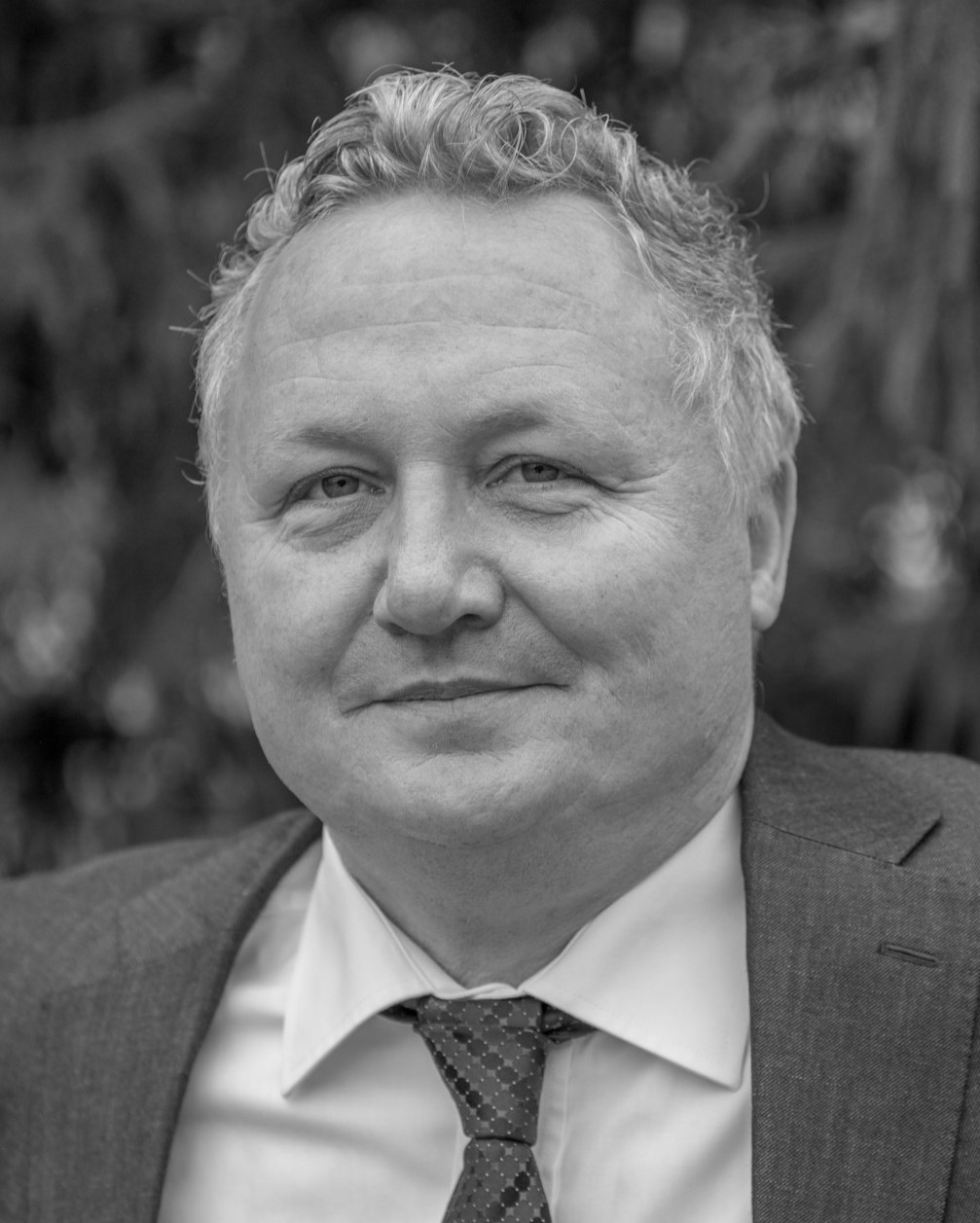}}]
{Thomas Fahringer} 
received his Ph.D. degree from the Vienna University of Technology in 1993. Since 2003, he has been a full professor of Computer Science at the Institute of Computer Science, University of Innsbruck, Austria. His main research interests include software architectures, programming paradigms, compiler technology, performance analysis, and prediction for parallel and distributed systems.
\end{IEEEbiography}

\begin{IEEEbiography}[{\includegraphics
[width=1in,height=1.25in,clip,
keepaspectratio]{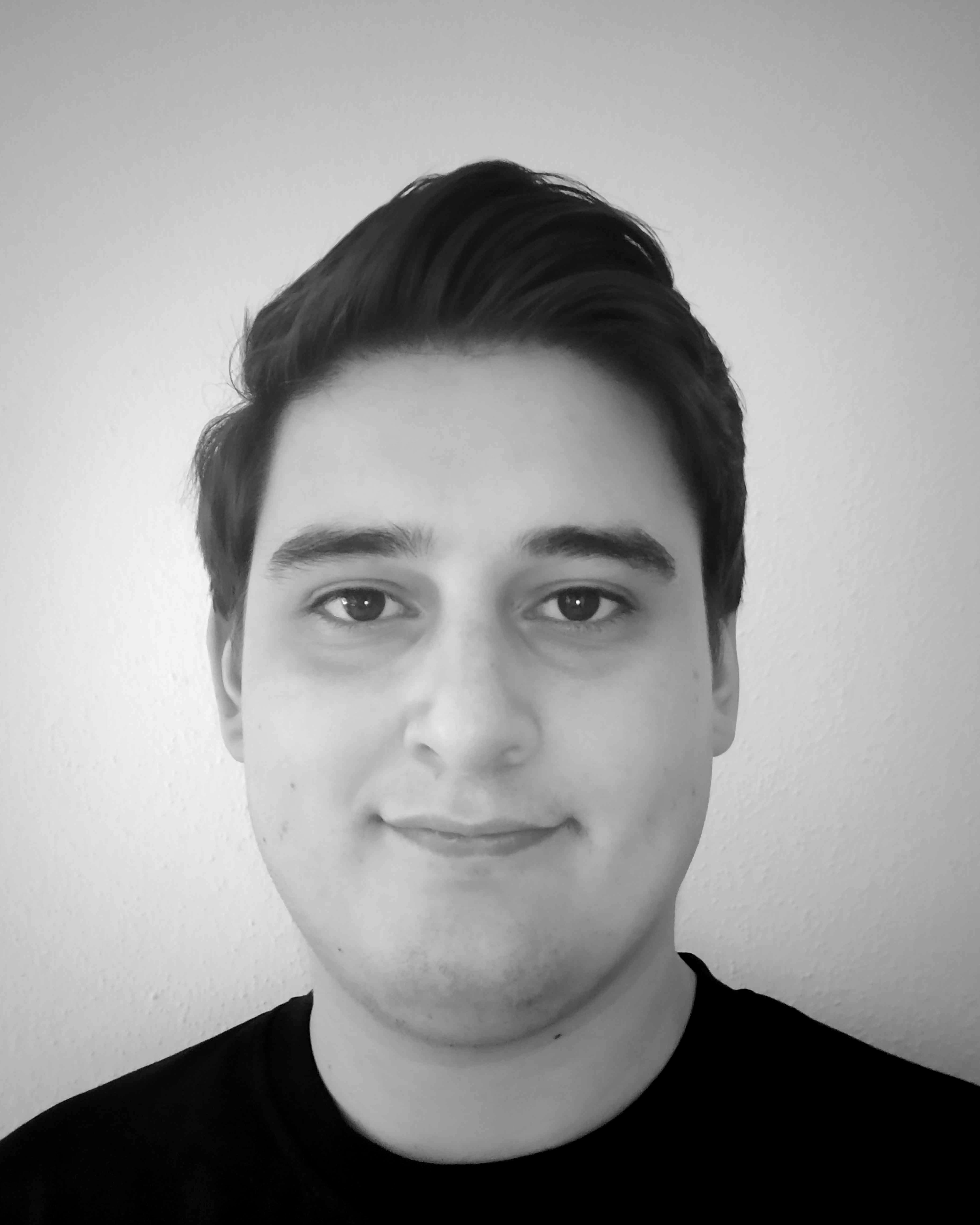}}]
{Felix Erlacher} 
received his B.Sc. in Computer Science from the University of Innsbruck, Austria, in 2021. He is currently working towards an M.Sc. in Computer Science. His research interests include distributed systems, programming paradigms, information security, and information retrieval.
\end{IEEEbiography}

\begin{IEEEbiography}[{\includegraphics
[width=1in,height=1.25in,clip,
keepaspectratio]{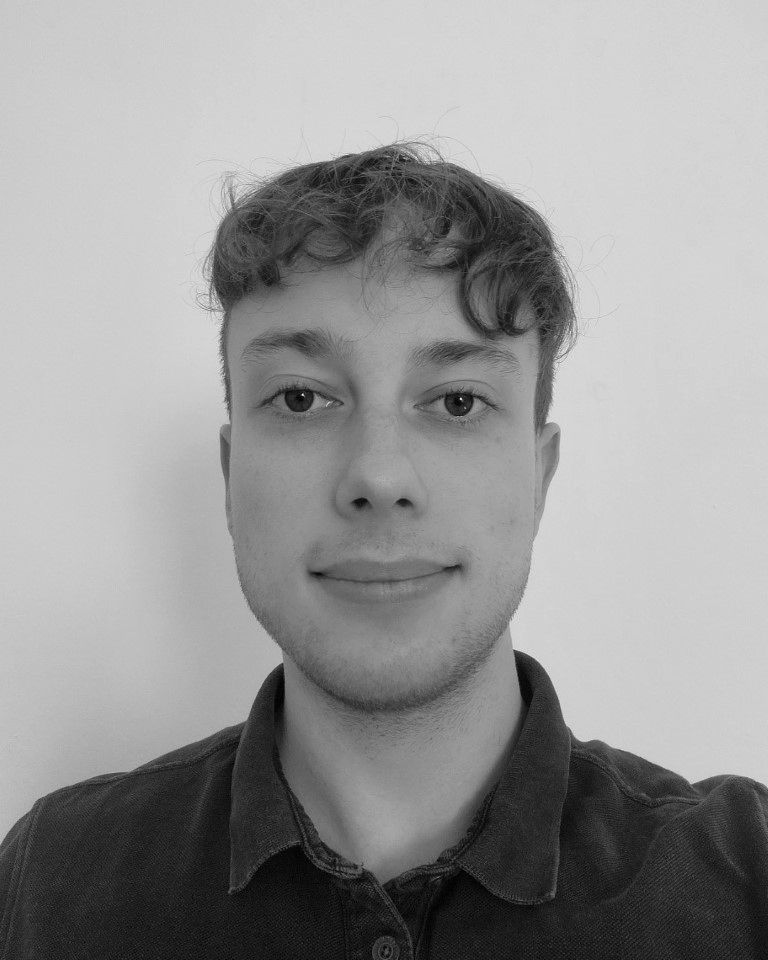}}]
{Elias Kohler} 
received his B.Sc. in Computer Science from the University of Innsbruck, Austria, in 2023. He is currently working towards an M.Sc. in Software Engineering. 
\end{IEEEbiography}

\begin{IEEEbiography}[{\includegraphics
[width=1in,height=1.25in,clip,
keepaspectratio]{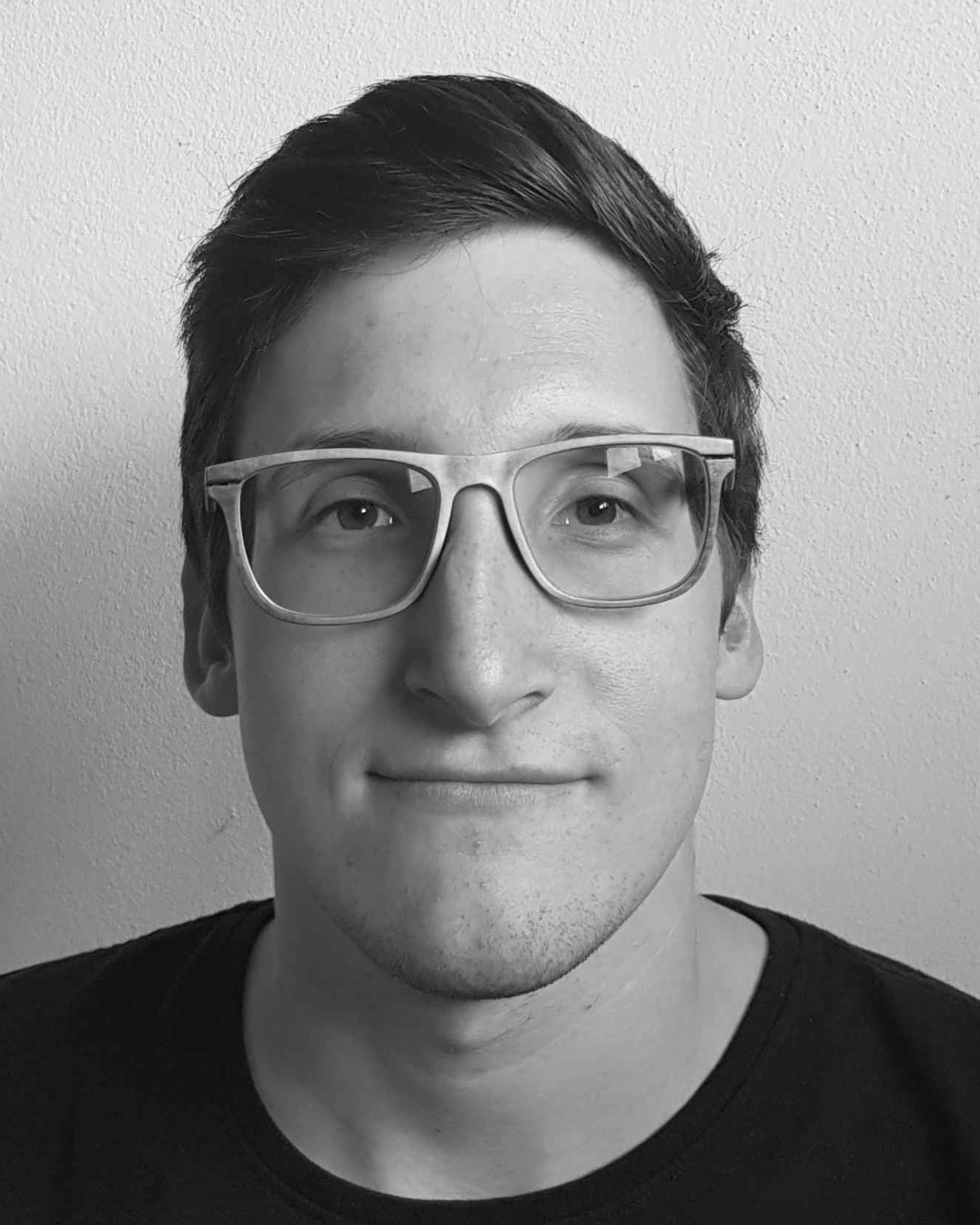}}]
{Stefan Pedratscher} 
received his M.Sc. in Computer Science from the University of Innsbruck, Austria, in 2021. He is a Ph.D. student in the Distributed and Parallel Systems Group at the University of Innsbruck, Austria, and his main research interests include serverless computing and stream processing in the Edge-Cloud continuum.
\end{IEEEbiography}

\begin{IEEEbiography}[{\includegraphics
[width=1in,height=1.25in,clip,
keepaspectratio]{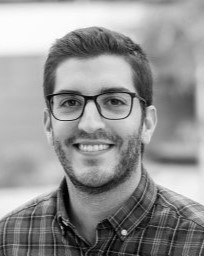}}]
{Juan Aznar-Poveda} 
received his Ph.D. degree in Telecommunications Engineering from the Technical University of Cartagena, Spain, in 2022. He is a postdoctoral researcher at the Distributed and Parallel Systems Group of the University of Innsbruck, Austria. His research interests include distributed systems, distributed databases, artificial intelligence, and wireless communications.
\end{IEEEbiography}

\begin{IEEEbiography}[{\includegraphics
[width=1in,height=1.25in,clip,
keepaspectratio]{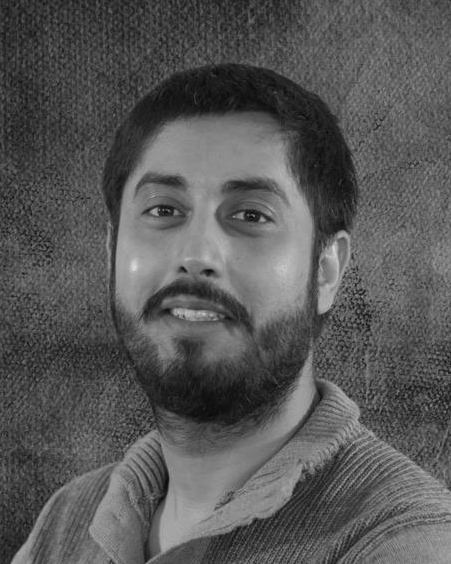}}]
{Nishant Saurabh} 
received his Ph.D. degree in computer science from the University of Innsbruck, Austria. He is currently an assistant professor at Utrecht University, The Netherlands. His research interests include Edge-Cloud computing distributed infrastructures and performance modeling.
\end{IEEEbiography}

\begin{IEEEbiography}[{\includegraphics
[width=1in,height=1.25in,clip,
keepaspectratio]{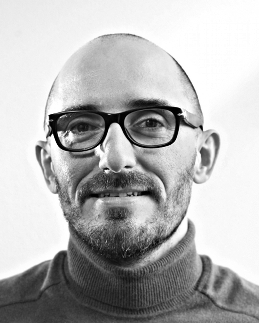}}]
{Adrien Lebre} 
received his PhD degree from the Grenoble Institute of Technologies, and the habilitation degree from the University of Nantes. He is a senior researcher at Inria, Nantes, France.
\end{IEEEbiography}

\end{document}